\documentclass[preprint, nobibnotes, aps, superscriptaddress]{revtex4-1}
\usepackage{amsmath}
\usepackage{amssymb}
\usepackage{hyperref}
\usepackage{physics}
\usepackage{graphicx}
\usepackage{color}

\begin{document}

\title{Designing Multi--Functional Metamaterials}

\author{J. R. Capers}
\email{jrc232@exeter.ac.uk}
\affiliation{Department of Physics and Astronomy, University of Exeter, Stocker Road, Exeter, EX4 4QL}

\author{S. J. Boyes}
\affiliation{DSTL, Porton Down, Salisbury, Wiltshire, SP4 0JQ}

\author{A. P. Hibbins}
\affiliation{Department of Physics and Astronomy, University of Exeter, Stocker Road, Exeter, EX4 4QL}

\author{S. A. R. Horsley}
\affiliation{Department of Physics and Astronomy, University of Exeter, Stocker Road, Exeter, EX4 4QL}

\date{\today}

\begin{abstract}
    The ability to design passive structures that perform different operations on different electromagnetic fields is key to many technologies, from beam--steering to optical computing.
    While many techniques have been developed to optimise structure to achieve specific functionality through inverse design, designing multi--function materials remains challenging.
    We present a semi--analytic method, based on the discrete dipole approximation, to design multi--functional metamaterials.
    To demonstrate the generality of our method, we design a device that operates at optical wavelengths and beams light into different directions depending on the source polarisation and a device that works at microwave wavelengths and sorts plane waves by their angle of incidence.
\end{abstract}

\maketitle

\section{Introduction}

Much of modern technology is enabled by the control of electromagnetic waves.
One way to manipulate electromagnetic radiation is to structure materials in space \cite{Munk2000, Leonhardt2006, Pendry2006}, and more recently time \cite{Galiffi2022, Ptitcyn2022}, to control electromagnetic fields.
Metamaterials, man--made materials that have electromagnetic properties associated with their structure rather than chemistry, have demonstrated extraordinary control over all types of wave \cite{Kadic2019}.
Typically metamaterials are built from `meta--atoms', sub--wavelength resonant scattering elements with scattering properties that depend upon their structure.
By tuning the geometry of the meta--atoms, almost any wave scattering effect can be realised.
Of particular interest is the ability to design multi--functional metamaterials: materials that scatter differently in response to different input fields.
Being able to passively perform different operations upon different waves is key to several problems across physics; from multiplexing output beams \cite{Pande2020} to mode sorting \cite{Frellsen2016, Piggott2015} and beam steering \cite{Li2019}.
Furthermore, metamaterials comprising several discrete meta-atoms that can be arranged anywhere in space, and that are explicitly aperiodic, are typically designed using genetic algorithms \cite{Wiecha2017}.
These are extremely effective at exploring large search spaces, but are numerically expensive and produce results that can be difficult to interpret \cite{Yeung2020}.

To understand why the design of multi--functional materials is challenging, we consider an electric field $\vec{\phi} (\boldsymbol{r})$ of a fixed frequency $\omega = c k_0$, with wave number $k_0$, in a material with a spatially varying permittivty $\varepsilon (\boldsymbol{r})$.
This wave obeys the vector Helmholtz equation,
\begin{equation}
    \nabla \times \nabla \times \vec{\phi} (\boldsymbol{r})  + k_0^2 \varepsilon (\boldsymbol{r}) \vec{\phi} (\boldsymbol{r}) = 0 .
    \label{eq:waveEqn}
\end{equation}
The difficulty in the design of multi--functional materials is the problem of finding a single material distribution (here the permittivity, $\varepsilon$) that performs two (or more) desired wave transformations.
This means that both of the desired wave behaviours, $\vec{\phi}_1 (\boldsymbol{r})$ and $\vec{\phi}_2 (\boldsymbol{r})$ must be solutions to the same Helmholtz equation,
\begin{align}
    \nabla \times \nabla \times \vec{\phi}_1 (\boldsymbol{r})  + k_0^2 \varepsilon (\boldsymbol{r}) \vec{\phi}_1 (\boldsymbol{r}) &= 0, & \nabla \times \nabla \times \vec{\phi}_2 (\boldsymbol{r})  + k_0^2 \varepsilon (\boldsymbol{r}) \vec{\phi}_2 (\boldsymbol{r}) = 0 .
\end{align}
From this statement, we can find a condition upon the two wave--fields for this to be possible.
Multiplying the first of these by $\vec{\phi}_2 (\boldsymbol{r})$ and the second by $\vec{\phi}_1 (\boldsymbol{r})$, then taking the difference eliminates the material properties such that
\begin{equation}
    \vec{\phi}_2 (\boldsymbol{r}) \cdot \left[ \nabla \times \nabla \times \vec{\phi}_1 (\boldsymbol{r}) \right] - \vec{\phi}_1 (\boldsymbol{r}) \cdot \left[ \nabla \times \nabla \times \vec{\phi}_2 (\boldsymbol{r}) \right] = 0.
\end{equation}
Integrating this over all space, then using Green's vector identity \cite{Stratton2007}, we find that 
\begin{equation}
    \oint_{\partial V} \left[ \vec{\phi}_2 (\boldsymbol{r}) \times \nabla \times \vec{\phi}_1 (\boldsymbol{r}) - \vec{\phi}_1 (\boldsymbol{r}) \times \nabla \times \vec{\phi}_2 (\boldsymbol{r}) \right] \cdot d\boldsymbol{S} = 0.
    \label{eq:condition_surface}
\end{equation}
This places a stringent condition on the two wave--fields if they are to be supported by the same material, which can be used to derive fundamental bounds on the performance of multi--functional devices \cite{Miller2021}.  Eq. (\ref{eq:condition_surface}) is a generalization of Poynting's theorem, representing the conservation of the norm of the system modes; ensuring for example, their orthogonality.  
To better understand the connection between the above and energy conservation, consider the special case where we demand the same permittivity distribution supports the solution $\vec{\phi}_1 (\boldsymbol{r}) = \boldsymbol{E}$ and its complex conjugate (time reverse) $\vec{\phi}_2 (\boldsymbol{r}) = \boldsymbol{E}^*$.  
The surface integral (\ref{eq:condition_surface}) can then be re--written using the divergence theorem, 
\begin{equation}
    \nabla \cdot \left( \boldsymbol{E} \times \nabla \times \boldsymbol{E}^* - \boldsymbol{E}^* \times \nabla \times \boldsymbol{E} \right) = 0.
\end{equation}
Applying Maxwell's equations to convert the curls into magnetic fields $\nabla \times \boldsymbol{E} = -i\omega\eta_0 \boldsymbol{H}$, where $\eta_0$ is the impedance of free space, we find 
\begin{equation}
    \nabla \cdot \left( \boldsymbol{E} \times \boldsymbol{H}^* + \boldsymbol{E}^* \times \boldsymbol{H} \right) = 0 ,
\end{equation}
which is the usual expression for energy conservation expressed in terms of the Poynting vector $\boldsymbol{S} = \frac{1}{2} {\rm Re} \left[ \boldsymbol{E} \times \boldsymbol{H}^* \right]$.

A direct application of Eq. (\ref{eq:condition_surface}) to design multi--functional materials is generally difficult.  
Instead several other design methodologies have emerged recently \cite{Molesky2018}.
Where the function of the metasurface, a 2D metamaterial, is to impart a known phase and amplitude offset, the Gerchberg--Saxton algorithm \cite{Gerchberg1972} is commonly used \cite{Liu2021}.
While simple and efficient this method can struggle to capture the coupling between elements of the metamaterial, requiring either strong field confinement within, or large spacing between the elements.
As well as this, several full--wave simulations are required to build up a library of phase and amplitude changing meta--atoms.
For problems where a continuous permittivity distribution is required, `topology optimisation' \cite{Bendsoe2003} can be used to design structures that sort waveguide modes \cite{Frellsen2016} or perform wavelength--dependant behaviour \cite{Piggott2015}.
This method typically uses a full--wave solver to find the fields then gradient descent optimisation to design a structure that optimises a figure of merit.
For large structures this can be computationally demanding, however the adjoint method can be used to optimise a figure of merit by converting a shape derivative for the entire design into two field calculations \cite{Lalau-Keraly2013}.
Even with this greatly improved numerical efficiency, many full--wave simulations are still required over the course of the optimisation.

We present a general framework to design multi-functional metamaterials in electromagnetism in order to overcome some of the aforementioned difficulties.
Based on the discrete dipole approximation \cite{Purcell1973}, the framework proposed in this paper is numerically efficient while still being easily applicable to a wide range of electromagnetics problems.
The discrete dipole approximation is briefly described in Section \ref{sec:modelling} then a summary of how metamaterials can be designed within this approximation is provided in in Section \ref{sec:design}.
Considering the problem of shaping the far--field of a dipole emitter while also increasing its efficiency, this is then extended  to design multi--functional materials in Section \ref{sec:multiobj}.
Several examples of utilising this framework are then shown in Section \ref{sec:examples}.
The first device we present operates at optical wavelengths $\lambda = 550$ nm, with silicon nanospheres used as the scattering elements, and beams light into different directions based upon the polarisaion of the source.
The second device we show works at microwave wavelengths $\sim 20$ mm (15.5 GHz), using `metacubes' \cite{Powell2020} as the scatterers, and sorts signals by their incidence direction.

\section{Modelling Metamaterials \label{sec:modelling}}

Before trying to design the scattering properties of a metamaterial, it is necessary to characterise the effect of the material upon an electromagnetic wave.
Maxwell's equations for a fixed frequency $\omega = ck_0$, where $k_0$ is the wave--number, can then be written as 
\begin{equation}
    \begin{pmatrix}
        \nabla \times \nabla \times & \boldsymbol{0} \\
        \boldsymbol{0} & \nabla \times \nabla \times
    \end{pmatrix}
    \begin{pmatrix}
        \boldsymbol{E} \\
        \boldsymbol{H}
    \end{pmatrix}
    - k_0^2
    \begin{pmatrix}
        \boldsymbol{E} \\
        \boldsymbol{H}
    \end{pmatrix}
    = 
    \begin{pmatrix}
        \boldsymbol{E}_{\rm inc} \\
        \boldsymbol{H}_{\rm inc}
    \end{pmatrix}
    + 
    \begin{pmatrix}
        \omega^2 \mu_0 & i \omega \mu_0 \nabla \times \\
        -i\omega \nabla \times & k_0^2
    \end{pmatrix}
    \begin{pmatrix}
        \boldsymbol{P} \\
        \boldsymbol{M}
    \end{pmatrix} 
    \label{eq:maxwell}
\end{equation}
where the material is characterised by the polarisation density $\boldsymbol{P}$ and magnetisation density $\boldsymbol{M}$ and the incident field is $(\boldsymbol{E}_{\rm inc}, \boldsymbol{H}_{\rm inc})^{T}$.
The incident field could be due to an emitter near the structure, or might be a plane wave.
Assuming that the scatterers are sub--wavelength in size means that the field can be treated as constant over the scatterer, and so the scatterer's distribution can be written as a delta--like point at its center $\boldsymbol{r}_n$.
The scatterer then acquires an electric and magnetic dipole moment in response to the applied fields
\begin{align}
    \boldsymbol{P} &= \sum_n \boldsymbol{\alpha}_E \boldsymbol{E} (\boldsymbol{r}_n) \delta (\boldsymbol{r}-\boldsymbol{r}_n)  & &\text{and} &
    \boldsymbol{M} &= \sum_n \boldsymbol{\alpha}_H \boldsymbol{H} (\boldsymbol{r}_n) \delta (\boldsymbol{r}-\boldsymbol{r}_n),
    \label{eq:PM}
\end{align}
where $\boldsymbol{\alpha}_E$ is the electric polarisability tensor and $\boldsymbol{\alpha}_H$ is the magnetic polarisability tensor.
The wave--equation with delta--like source terms can be solved with the appropriate Green's function \cite{Schwinger1950, Tai1993} as
\begin{equation}
    \phi (\boldsymbol{r}) = \phi_{\rm inc} (\boldsymbol{r}) + \sum_{n=0}^N G(\boldsymbol{r}, \boldsymbol{r}_n) \phi (\boldsymbol{r}_n) .
    \label{eq:waveSln}
\end{equation}
We have adopted the compact notation 
\begin{align}
    \phi (\boldsymbol{r}) &= 
    \begin{pmatrix}
    \boldsymbol{E} (\boldsymbol{r}) \\
    \eta_0 \boldsymbol{H} (\boldsymbol{r})
    \end{pmatrix} \ \text{and} &
    G(\boldsymbol{r}, \boldsymbol{r'}) &= 
    \begin{pmatrix}
        \xi^2 \boldsymbol{G}(\boldsymbol{r}, \boldsymbol{r'}) \boldsymbol{\alpha}_E & i \xi \nabla \times \boldsymbol{G}(\boldsymbol{r}, \boldsymbol{r'}) \boldsymbol{\alpha}_H \\ 
        -i\xi \nabla \times \boldsymbol{G}(\boldsymbol{r}, \boldsymbol{r'}) \boldsymbol{\alpha}_E & \xi^2 \boldsymbol{G}(\boldsymbol{r}, \boldsymbol{r'}) \boldsymbol{\alpha}_H
    \end{pmatrix} ,
\end{align}
where 
\begin{equation}
    \boldsymbol{G} (\boldsymbol{r}, \boldsymbol{r'}) = \left[ \boldsymbol{1} + \frac{1}{\xi^2} \nabla \otimes \nabla \right] \frac{e^{i\xi|\boldsymbol{r} - \boldsymbol{r'}|}}{4 \pi |\boldsymbol{r} - \boldsymbol{r'}|}
\end{equation}
is the Dyadic Green's function and $\xi$ is a dimensionless wave--number.
To completely specify the field solution (\ref{eq:waveSln}), the fields applied to each scatterer $\phi (\boldsymbol{r}_n)$ must be determined.
This includes the source field, as well as all orders of multiple scattering interactions between the scatterers themselves.
The applied fields $\phi (\boldsymbol{r}_n)$ can be found through the self--consistency condition \cite{Foldy1945} 
\begin{equation}
    \boldsymbol{R}_{mn} \boldsymbol{\phi}_m = \boldsymbol{\phi}_{i,n} ,
\end{equation}
with $\boldsymbol{R}_{nm} = \boldsymbol{1}\delta_{nm} - G(\boldsymbol{r}_n, \boldsymbol{r}_m)$, $\boldsymbol{\phi}_m = \phi(\boldsymbol{r}_m)$ and $\boldsymbol{\phi}_{i,n} = \phi_i(\boldsymbol{r}_n)$.
This forms a linear system that can be solved for the fields applied to the scatterers $\boldsymbol{\phi}_m$ using standard matrix methods.
Once these are found, the fields (\ref{eq:waveSln}) are fully specified.

\section{Designing Metamaterials \label{sec:design}}

The scattering properties of a metamaterial made of several sub--wavelength scatterers can be designed using the methodology presented in \cite{Capers2021}.
We briefly review this here, before extending the method to multi--functional devices. By
Taylor expanding the delta function sources in the wave equation (\ref{eq:PM}) under small changes in the position of a scatterer yields,
\begin{equation}
    \delta (\boldsymbol{r} - \boldsymbol{r}_n - \delta \boldsymbol{r}_n) = \delta (\boldsymbol{r} - \boldsymbol{r}_n) - (\delta \boldsymbol{r}_n \cdot \nabla ) \delta (\boldsymbol{r} - \boldsymbol{r}_n) + \frac{1}{2} (\delta \boldsymbol{r}_n \cdot \nabla )^2 \delta (\boldsymbol{r} - \boldsymbol{r}_n) + \ldots ,
    \label{eq:fieldVar}
\end{equation}
and by keeping only terms linear in $\delta \boldsymbol{r}_n$ an expression for how the field $\phi(\boldsymbol{r})$ changes due to a small change in the location of scatterer $n$ can be deduced, 
\begin{equation}
    \delta \phi (\boldsymbol{r}) = G(\boldsymbol{r}, \boldsymbol{r}_n) \nabla \phi (\boldsymbol{r}_n) \cdot \delta \boldsymbol{r}_n .
    \label{eq:field_var}
\end{equation}
A figure of merit $\mathcal{F}$ can be written as a functional of the field $\mathcal{F} = \mathcal{F}[\phi]$.
Under small changes in the field as a result of a small change in the position of a scatterer, the figure of merit is changed by a small amount.
Given a particular figure of merit, an analytic expression for this change $\delta \mathcal{F} [\phi, \delta \phi]$ can be derived.
Since this will be linear in the change in position of a scatterer $\delta \boldsymbol{r}_n$, the resulting expression allows for the derivative of the figure of merit with respect to the scatterer locations $\pdv{\mathcal{F}}{\boldsymbol{r}_n}$ to be calculated analytically.
As an example, say the figure of merit is the amplitude squared of the field at a particular location $|\phi(\boldsymbol{r}_\star)|^2$.
Expanding under small changes in the field we can find how the figure of merit changes,
\begin{align}
    \mathcal{F} &= |\phi(\boldsymbol{r}_\star)|^2 , \\
    \delta \mathcal{F} &= 2 {\rm Re} \left[ \phi^*(\boldsymbol{r}_\star) \delta \phi(\boldsymbol{r}_\star) \right] .
\end{align}
Substituting into this in the variation of the field (\ref{eq:fieldVar}) and diving by $\delta \boldsymbol{r}_n$ we find that 
\begin{align}
    \pdv{\mathcal{F}}{\boldsymbol{r}_n} &= 2 {\rm Re} \left[ \phi^*(\boldsymbol{r}_\star) G(\boldsymbol{r}_\star, \boldsymbol{r}_n) \nabla \phi (\boldsymbol{r}_n) \right] .
\end{align}
This approach gives an analytic expression for the derivative of a figure of merit that can be evaluated for all of the scatterers at the same time then used in a gradient descent optimisation \cite{Shwartz2014}
\begin{equation}
    \boldsymbol{r}_n^{i+1} = \boldsymbol{r}_n^{i} + \gamma \pdv{\mathcal{F}}{\boldsymbol{r}_n} ,
    \label{eq:gradient_descent}
\end{equation}
where $\gamma$ is the learning rate and $i$ is the iteration number.
While the derivatives of the fields still need to be evaluated, the  derivative of the figure of merit, which is more numerically expensive, does not.

Extending this to apply to multi--functional metamaterials, requires one to seek to increase some set of figures of merit $\left\{ F_1,F_2, F_3 \ldots \right\}$.
A composite figure of merit can be constructed that is a weighted sum of these,
\begin{equation}
    \mathcal{F} = \sum_i w_i F_i,
\end{equation}
where $w_i$ are the weights for each figure of merit. 
This composite figure of merit can be optimised in the same way as a single figure of merit, where the gradient in equation (\ref{eq:gradient_descent}) is now
\begin{equation}
    \pdv{ \mathcal{F} }{\boldsymbol{r}_n} = \sum_i w_i \pdv{ F_i }{\boldsymbol{r}_n} .
    \label{eq:multiFgrad}
\end{equation}
Key to the success of this method is a sensible choice of the weights, $w_i$.
An appropriate choice can be informed by considering the desired properties of the resulting device, as we explore in the next section.

\section{Multi--Objective Optimisation Considerations \label{sec:multiobj}}

\begin{figure}
    \centering
    \includegraphics[width=0.8\linewidth]{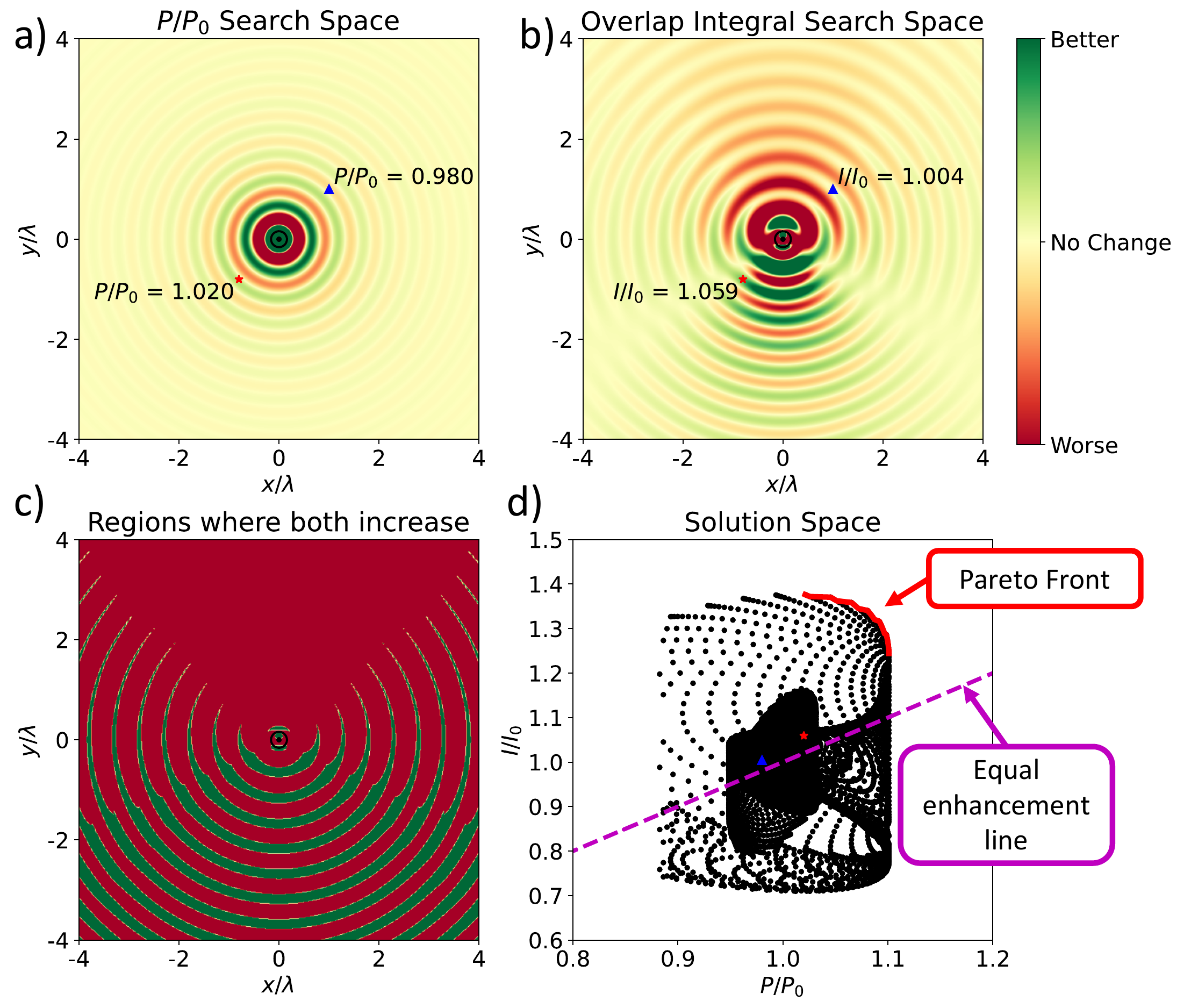}
    \caption{An example of a multi--objective optimisation problem. 
    We seek to shape the radiation pattern of a dipole emitter pointing out of the plane along the $z$ axis, while also increasing its efficiency.
    To achieve this, we consider placing a single isotropic scatterer near the emitter and explore the effect this has upon the figures of merit; the power emission and the radiation pattern.
    a) Shows how placing a scatterer at a particular point changes the power emission and b) shows how the overlap integral is changed.  
    These represent the search spaces of the problem.  
    The additional difficulty posed by \emph{multi}, rather than single, objective problems is shown in c).  
    A scatterer can only be placed in the green regions, where both figures of merit are enhanced.
    Each point in the search spaces correspond to a point in solution space, shown in d).  
    The blue triangle and red star and shown in both search and solution spaces to demonstrate this.
    Also shown in the solution space, d), is the diagonal `line of equal enhancement' representing equal performance of each figure of merit and the Pareto front \cite{Hwang1979}, which represents the acceptable solutions to the multi--objective problem.
    }
    \label{fig:fig1}
\end{figure}
Consider a simple example problem, shown in Figure \ref{fig:fig1}.
The goal is to distribute scatterers around a point emitter at location $\boldsymbol{r'}$ with polarisation $\boldsymbol{p} = \boldsymbol{\hat{z}}$ such that \emph{two} figures of merit are simultaneously maximised.
More specifically, the goal is  to re--shape the radiation pattern of an emitter, while simultaneously increasing the efficiency.
The first figure of merit is the power emission of the emitter,
\begin{equation}
    F_1 = P = \frac{\omega}{2} {\rm Im}[\boldsymbol{p}^* \cdot \boldsymbol{E} (\boldsymbol{r'})] .
    \label{eq:fom_pwr}
\end{equation}
The second is the overlap integral between the angular distribution of the Poynting vector in the far--field, $|\boldsymbol{S} (\theta)|$ and the desired angular distribution $\psi_T (\theta)$
\begin{equation}
    F_2 = I = \frac{ \int d\theta |\boldsymbol{S} (\theta)| \psi_T (\theta) }{\sqrt{\int d\theta |\boldsymbol{S}(\theta)|^2} \sqrt{\int d\theta \psi_T^2 (\theta)}} ,
    \label{eq:fom_overlap}
\end{equation}
where the angle $\theta$ is in the same plane as the metasurface.
In the following examples,the target distribution is
\begin{equation}
    \psi_T (\theta) =
    \begin{cases}
        \cos^2 \theta \ 270^\circ < \theta < 90^\circ , \\
        0 \ \text{otherwise} .
    \end{cases}
\end{equation}
Both of these can be expanded to first order to find the gradient of the figure of merit with respect to the scatterer locations \cite{Capers2021}.
These expansions are given in Appendix A.
For convenience, both will be normalised by their free--space values, $P_0$ and $I_0$: these are the values of the figures of merit without any scatterers present.
Considering the effect of a single scatterer upon these figures of merit, Figure \ref{fig:fig1} a) and b) show how placing the scatterer in a particular location increases or decreases each figure of merit.
These maps define the search space for the problem.  
For multi--functional problems, however, there is the additional constraint that a scatterer should only be placed where \emph{both} figures are merit are increased.
This is shown in Figure \ref{fig:fig1}c; it is clear that multi--functional problems are significantly constrained and have complex search spaces.
Each scatterer location in the search spaces Figure \ref{fig:fig1} a) and b) corresponds to a point in the solution space, shown in Figure \ref{fig:fig1}d.
To demonstrate this correspondence, the red star and blue triangle are shown in the search and solution spaces.
Each point in solution space corresponds to a configuration of scatterers, which in our example is one, but could be any number.
The solution space, Figure \ref{fig:fig1}d, has a few interesting features.
The Pareto front \cite{Hwang1979}, shown as a red line, are all acceptable solutions to the multi--objective optimisation problem (i.e. where one figure of merit cannot be improved without sacrificing the other).
Along the diagonal, the dashed magenta line, enhancement of the two figures of merit is equal, which is often the desired outcome.
It would not be very useful to select a solution point where the power emission is large but the overlap is small, even if it lies on the Pareto front.
This observation informs the choice of weights in the optimisation procedure.
The weights are chosen to be proportional to the figure of merit itself, 
\begin{align}
    w_i \propto \frac{1}{F_i} \ \text{, and normalised so that} \sum_i w_i = 1 .
    \label{eq:weights}
\end{align}
Choosing the weights to be proportional to $1/F_i$ means that when the figure of merit is small the contribution of the gradient associated with that figure of merit to the sum (\ref{eq:multiFgrad}) is large, but when the figure of merit is large the contribution is suppressed.
Note that the figures of merit must be normalised so that their magnitudes can be meaningfully compared, for example by division by a free space value. 
Choosing the weights in this way allows for the design of multi--functional metamaterials built from discrete scatterers, for a variety of applications.
A few examples are offered in the following section.

\section{Multi--functional Devices \label{sec:examples}}

\begin{figure}
    \centering
    \includegraphics[width=0.8\linewidth]{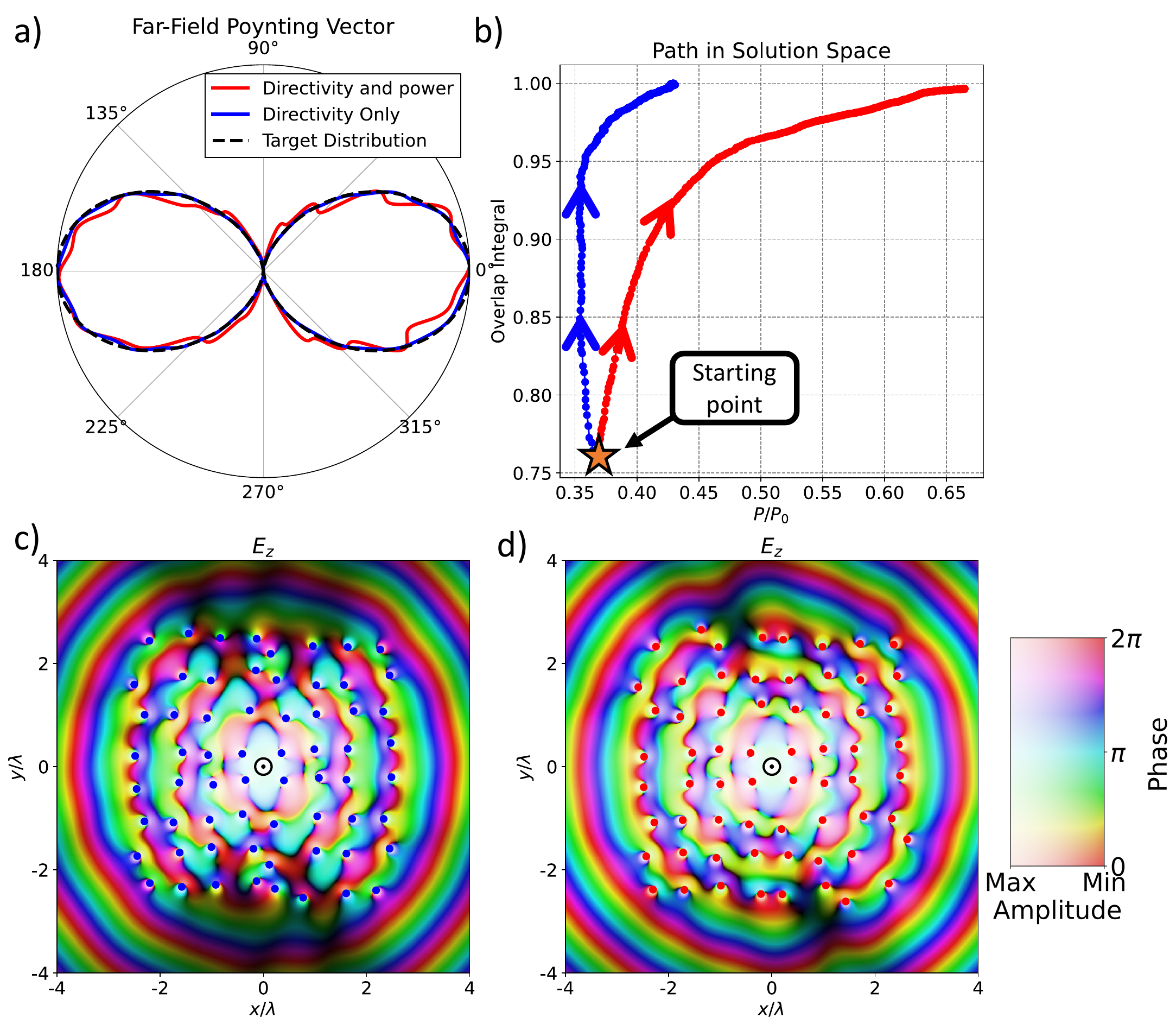}
    \caption{Multi--objective optimisation solutions, seeking to increase the emitted power from a dipole while also shaping the far--field radiation pattern into the desired double--lobed shape.
    For comparison, the single--objective case, where only the radiation pattern is shaped is shown.
    a) shows the far--field radiation pattern in the plane of the multi--functional structure (red) as well as the target radiation pattern (black dashes). 
    The case where only radiation pattern is shaped is shown in blue.
    In b), the paths in solution space of the single and multi--objective cases are shown.
    It is clear that our choice of weightings works well: both figures of merit undergo similar enhancements from their starting values.
    When only radiation pattern is controlled (blue line) power emission changes little over the optimisation, however when it is part of the composite figure of merit (red line) clear enhancement is seen at the same time as the overlap integral is increased.
    The single--objective structure is shown in c) and the multi--objective structure is show in d), with the emitter polarised out of the page at the origin.  
    In this example, we work at $\lambda  = 550$ nm and the scatterers are silicon spheres of radius 65 nm.}
    \label{fig:fig2}
\end{figure}

Continuing with the example developed in the previous section and shown in Figure \ref{fig:fig1}, the multi--objective problem of designing a structure that re--shapes the radiation pattern of an emitter in the plane of the metasurface, while also increasing efficiency, is addressed.
We work at a wavelength of $\lambda = 550$ nm and the scatterers are small silicon spheres of radius 65 nm.
For this system, the polarisability tensor can be found analytically from the Mie $a_1$ and $b_1$ coefficients.
Our figure of merits are the power emission (\ref{eq:fom_pwr}) and the overlap with the desired radiation pattern (\ref{eq:fom_overlap}) and we choose the weights according to (\ref{eq:weights}).
Analytic expression for the gradients of these figures are merit can be found analytically.
Using these gradients, weights (\ref{eq:weights}) and the gradient descent method (\ref{eq:gradient_descent}), we design the structures shown in Figure \ref{fig:fig2}.
For comparison, we also consider the single--objective case where only the far--field radiation pattern is shaped.
The resulting far--field radiation patterns are shown in Figure \ref{fig:fig2}a, with the path of the optimisation in solution space shown in Figure \ref{fig:fig2}b and the two resulting structures shown in Figure \ref{fig:fig2}c and d.
Examining first the radiation pattern, we note that the multi--objective optimisation produces a slightly worse match to the target distribution than the single--objective case.
This is due to the trade--off between the two figures of merit we seek to optimise.
In solution space, Figure \ref{fig:fig2}b, we see that in the case where only the radiation pattern is shaped (blue line) only a very small change in emitted power is seen.
Conversely, when both power and radiation pattern are optimised, the emitted power approximately doubles while the overlap integral also increases.
Unlike the case for a single scatterer shown in Figure \ref{fig:fig1}, it is impossible to plot the whole search space and determine the location of the Pareto front.
The scatterers have a diameter $\sim \lambda / 4$ and our solution box has size $8 \lambda$, meaning that there are $32^2 = 1024$ possible `pixels' a scatterer could occupy.
This means that for $N$ scatterers, the number of possible solutions is $1024!/(N! (1024-N)!)$.  
For $N = 64$, this is $\approx 10^{21}$.
Despite the exceedingly large search space, which even a genetic algorithm would explore only a very small portion of, our method has found a solution that performs well.
A comparison between solving this problem using the method we present and a genetic algorithm is given in Appendix B.

The second example we consider is manipulating the radiation pattern based on source polarisation.
Again, we work at $\lambda = 550$ nm and use 65 nm silicon spheres as the scatterers.
We aim to create beams at angles $\theta_i$, associated with source polarisation $\boldsymbol{p}_i$.
Our figures of merit are therefore
\begin{equation}
    F_i = |\boldsymbol{S} (\theta_i)| .
\end{equation}
The expansion of this to find analytically the gradient is given in Appendix A.
We consider the source polarisation being either left or right circularly polarised, i.e.
\begin{equation}
    \boldsymbol{p} = \frac{1}{\sqrt{2}}
    \begin{pmatrix}
        1 \\
        \pm i \\
        0
    \end{pmatrix} .
\end{equation}
The Poynting vector can then be expanded to first order to find the derivatives of the figures of merit for the optimisation procedure.
\begin{figure}
    \centering
    \includegraphics[width=0.8\linewidth]{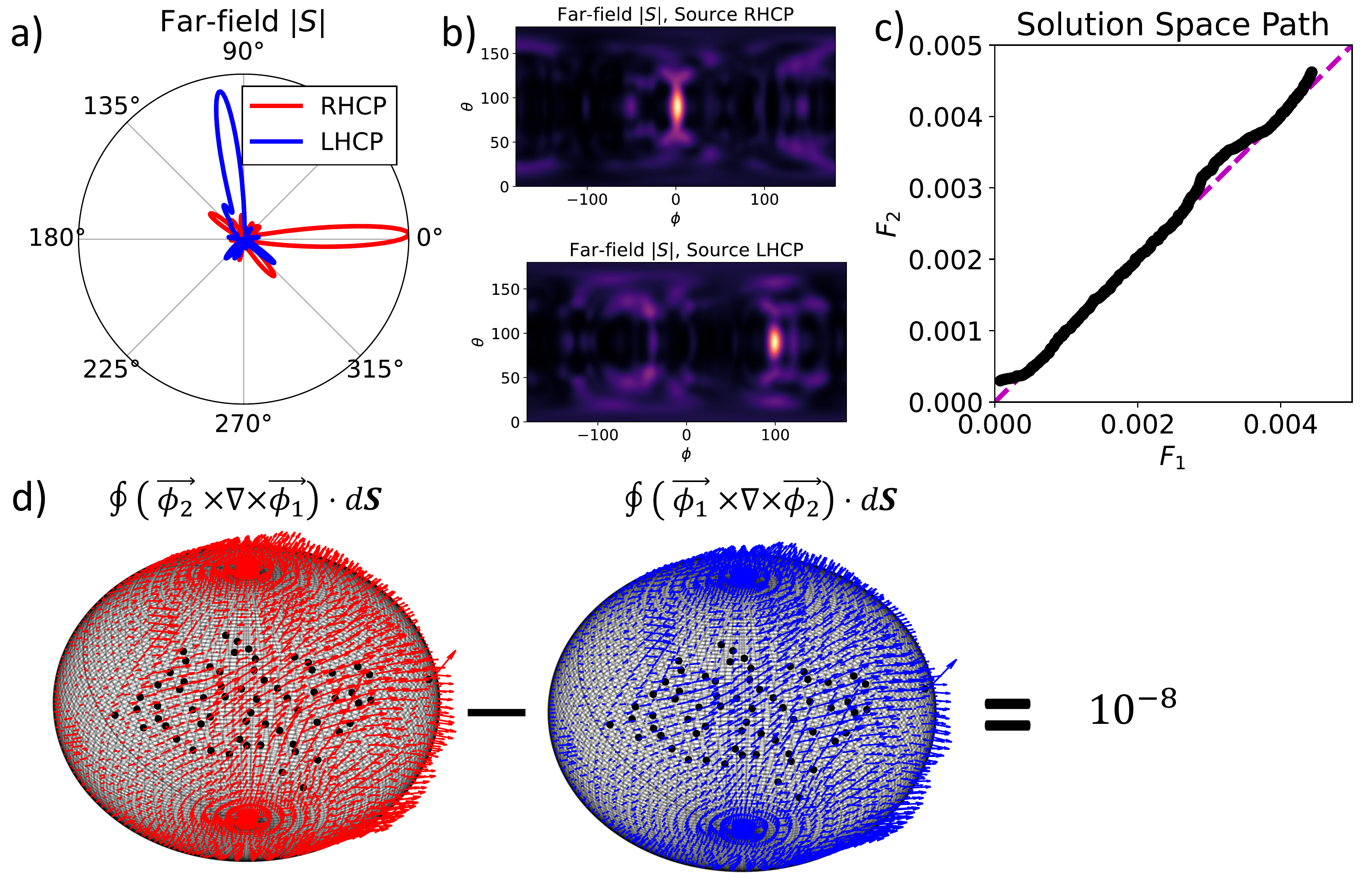}
    \caption{Solution to the multi--objective problem of beaming in different directions based on the polarisation of the source.
    We work at optical wavelengths $\lambda = 550$ nm, using silicon spheres of radius $65$ nm as the scatterers.
    The far--field Poynting vector is shown a) in the plane of the metamaterial and b) the full far--field sphere defined by the in--plane angle $\theta$ and the polar angle $\phi$ for left and right handed circularly polarised sources.
    The aim was for a right handed source to beam into the 0$^\circ$ direction and for a left handed source to beam into the 100$^\circ$ direction.
    The path in solution space of the optimisation, c), shows that over the optimisation both figures of merit are enhanced equally, due to our choice of weights.
    The multi--functionality condition (\ref{eq:condition_surface}) is verified in panel d), by computing the integrals numerically.
    This yields 10$^{-8}$, which is small enough to be considered zero within the numerical error associated with evaluating the surface integral.}
    \label{fig:fig3}
\end{figure}
Figure \ref{fig:fig3}a shows the radiation patterns of the designed structure excited by each of the two different sources we consider.
For a right-handed source, the target angle is $\theta = 0^\circ$ and for a left--handed source, $\theta = 100^\circ$.
The far--field Poynting vector, Figure \ref{fig:fig3}b, also shows clear peaks at the desired locations.
The path in solution space, Figure \ref{fig:fig3}c, shows that the choice of weighting has ensured that the performance of both figures of merit remain similar over the optimisation and in the final result. 
The multi--functionality condition (\ref{eq:condition_surface}) is considered in Figure \ref{fig:fig3}d.
Forming the vector fields $\vec{\phi}_i \times \nabla \times \vec{\phi}_j$ on the surface of a sphere enclosing the structure, integrating over the surface and taking the difference yields a result of the order $10^{-8}$, within expected numerical error.

\begin{figure}
    \centering
    \includegraphics[width=0.8\linewidth]{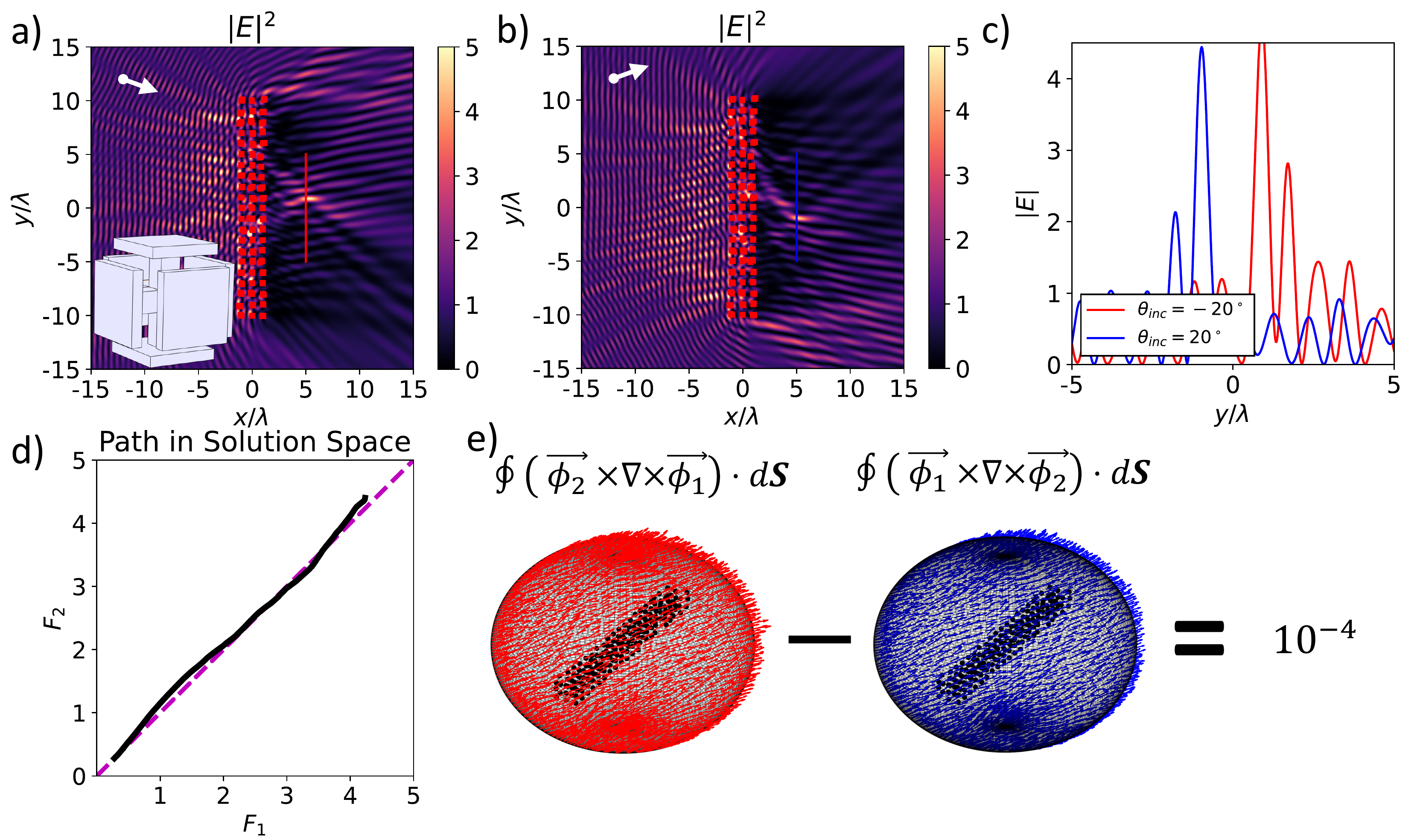}
    \caption{Design of a multi--functional device, with its operation depending upon the direction of incidence of a plane wave.
    If the wave is incident from $\pm 20^\circ$, the wave is focused to different locations.  
    The performance of the device under $- 20^\circ$ incident is shown in a) and under $20^\circ$ incidence in b), with direction of incidence shown as a white arrow.
    We work at 15.5 GHz using metacubes, shown inset in a), as the scatterers.
    Panel c), shows cuts of the fields under the two different incidence angles, demonstrating peaks at the target positions.
    The path in solution space, d), shows that the two figures of merit progress at the same rate over the optimisation, leading to a structure with roughly equal performance for each figure of merit.
    The multi--functionality condition (\ref{eq:condition_surface}) is verified in e) by evaluating the surface integrals.  
    In this case, the result is $10^{-4}$.  
    The numerical error here is larger than the results in Figure \ref{fig:fig3} due to the strongly oscillatory nature of the integrand, making evaluation of the surface integral more sensitive.}
    \label{fig:fig4}
\end{figure}

The third and final example we consider is designing a device for beam sorting.
Working at 15.5 GHz and using `metacubes' \cite{Powell2020} as the scattering element.
The metacubes, formed of six metal faces joined by three connecting spokes, exhibit a strong dipole resonance at 15.5 GHz.  
Due to their complexity the polarisability tensor cannot be found analytically.
Instead, one can model a single scatterer under plane--wave incidence using a full--wave solver such as COMSOL \cite{comsol} and integrate over the currents to find the electric and magnetic dipole moments \cite{Arango2013, Liu2016} that may be converted to polarisability tensors.
Optimisation of a structure of many complex scatterers using such full--wave methods quickly becomes intractable.  
Our method presents the key benefit of being able to model large systems of potentially complicated scatterers, provided they can be approximated as dipolar, although it is possible to include higher order multipoles into the formalism \cite{Raab2005, Evlyukhin2011, Evlyukhin2013}.
The validity of using the discrete dipole approximation to describe these systems is verified by comparison with full--wave solution in Appendix C.
We seek a structure of metacubes that takes plane waves from different directions and focuses them to distinct points.
A device of this sort could be used, for example, to detect from which direction a signal is coming.
The figures of merit for this problem are, 
\begin{align}
    F_1 &= |\boldsymbol{E}_1 (\boldsymbol{r}_1)|\ \text{, and} & F_2 &= |\boldsymbol{E}_2 (\boldsymbol{r}_2)|,
\end{align}
where $\boldsymbol{r}_i$ denotes the location to focus the wave at for incident direction $i$ and $\boldsymbol{E}_i$ is the electric field produced by the structure under incidence from direction $i$.
The gradients of these figures of merit are given in Appendix A.
The structure resulting from this optimisation is shown in Figure \ref{fig:fig4}.
Operation of the device when driven by a TE plane wave incident at $20^\circ$ is shown in Figure \ref{fig:fig4}a and for a plane wave at $-20^\circ$ in Figure \ref{fig:fig4}b.
The two different focus points are clearly visible in the fields.
The path in solution space is shown in Figure \ref{fig:fig4}c, where again the choice of weighting has ensured roughly equal performance of $F_1$ and $F_2$.
Slices of the fields from Figure \ref{fig:fig4}a,b are shown in Figure \ref{fig:fig4}d, indicating the large main peaks at the desired focus locations. 
The validity of the multi--functionality condition is shown in Figure \ref{fig:fig4}e.
Here, the numerical error is larger due to the highly oscillatory nature of the integrands.

\section{Conclusions \& Outlook}

Beginning from general considerations of vector fields, we have derived a condition that the fields must satisfy if they are to be supported by the same material distribution.
Interestingly, this can be expressed as a surface integral so only the fields on a boundary are needed to determine whether a particular multi--functional device is feasible. 
While interesting, this condition is difficult to solve in general so instead we present an efficient and versatile semi--analytic method for designing multi--functional metamaterials made from discrete scattering elements.
Demonstrating the generality of our formulation, we apply our method to design multi--functional devices at both optical and microwave wavelengths.
We show structures that: i) enhance the efficiency of an emitter while shaping its radiation pattern; ii) beam in different directions based on the source polarisation; and iii) sort waves by their incidence direction.
In addition to the design method being simple, the structures we design are easy to fabricate as permittivity does not need to be graded.
Instead many identical resonators must be distributed in space.

Our approach could be utilised to design very wide classes of multi--functional devices. 
For example, the inclusion of higher--order multipoles could allow more complex resonators to be used, proving more degrees of freedom.
Due to the generality of the two central ideas behind our framework: expanding figures of merit analytically to avoid expensive numerical derivatives and the formulation of the multi--objective problem in Section \ref{sec:multiobj}, a wide range of devices could be designed using this approach.
Graded index structures as well as propagation in waveguides and fibre optical cables could all be engineered to depend upon polarisation or wavelength for communications, cloaking or sensing applications using the methodology we propose here.

\section*{Acknowledgements}

J.R.C would like to thank Dean Patient for many useful discussions and Alex Powell for providing the COMSOL model for the metacubes.

We acknowledge financial support from the Engineering and Physical Sciences Research Council (EPSRC) of the United Kingdom, via the EPSRC Centre for Doctoral Training in Metamaterials (Grant No. EP/L015331/1). 
J.R.C also wishes to acknowledge financial support from Defence Science Technology Laboratory (DSTL). 
S.A.R.H acknowledges financial support from the Royal Society (RPG-2016-186).
All data and code created during this research are openly available from the corresponding author, upon reasonable request.

\section*{Author Contributions}

J. R. C. and S.A.R.H conceived the idea. 
J.R.C derived the analytic expressions, performed the numerical simulations and wrote the manuscript. 
S.A.R.H, A.P.H. and S.J.B supervised the project. 
All authors commented on the manuscript.

\section*{Competing Interests}

The authors declare no competing interests.

\appendix

\section{Figure of merit expansions}

In this section we give expansions of the figures of merit used in the main text and the gradients that result from these expansions.
To write these, we will need to use the expressions for the changes in the fields that are given by (\ref{eq:field_var}) in the main text.
Unpacking the notation, we have,
\begin{align}
    \delta \boldsymbol{E} (\boldsymbol{r}) &= \left[ \xi^2 \boldsymbol{G}(\boldsymbol{r}, \boldsymbol{r}_n) \boldsymbol{\alpha}_E \nabla \boldsymbol{E} (\boldsymbol{r}_n) + i \xi \nabla \times \boldsymbol{G}(\boldsymbol{r}, \boldsymbol{r}_n) \boldsymbol{\alpha}_H \nabla \boldsymbol{H} (\boldsymbol{r}_n) \right] \delta \boldsymbol{r}_n , \label{eq:deltaE}   \text{and}\\
    \delta \boldsymbol{H} (\boldsymbol{r}) &= \left[ \xi^2 \boldsymbol{G}(\boldsymbol{r}, \boldsymbol{r}_n) \boldsymbol{\alpha}_H \nabla \boldsymbol{H} (\boldsymbol{r}_n) - i \xi \nabla \times \boldsymbol{G}(\boldsymbol{r}, \boldsymbol{r}_n) \boldsymbol{\alpha}_E \nabla \boldsymbol{E} (\boldsymbol{r}_n) \right] \delta \boldsymbol{r}_n \label{eq:deltaH}.
\end{align}

\subsection{Emitted power}

Beginning from the usual expression for emitted power, expanding under small changes in the fields and then substituting (\ref{eq:deltaE}) gives the gradient of the power emission with respect to scatterer positions analytically,
\begin{align}
    P &= \frac{\omega}{2} {\rm Im} \left[ \boldsymbol{p}^* \cdot \boldsymbol{E} (\boldsymbol{r'}) \right], \\
    \delta P &= \frac{\omega}{2} {\rm Im} \left[ \boldsymbol{p}^* \cdot \delta \boldsymbol{E} (\boldsymbol{r'}) \right],  \text{and}\\
    \nabla_{\boldsymbol{r}_n} P &= \frac{\omega}{2} {\rm Im} \left[ \boldsymbol{p}^* \cdot \left( \xi^2 \boldsymbol{G}(\boldsymbol{r'}, \boldsymbol{r}_n) \boldsymbol{\alpha}_E \nabla \boldsymbol{E} (\boldsymbol{r}_n) + i \xi \nabla \times \boldsymbol{G}(\boldsymbol{r'}, \boldsymbol{r}_n) \boldsymbol{\alpha}_H \nabla \boldsymbol{H} (\boldsymbol{r}_n) \right)\right] .
\end{align}

\subsection{Overlap Integral}

The figure of merit used to manipulate the shape of the Poynting vector in the far--field is the normalised overlap integral between the current angular distribution of the Poynting vector $|\boldsymbol{S} (\theta)|$ and the target angular distribution $\psi_T (\theta)$
\begin{equation}
    I = \frac{ \int d\theta |\boldsymbol{S} (\theta)| \psi_T (\theta) }{\sqrt{\int d\theta |\boldsymbol{S}(\theta)|^2} \sqrt{\int d\theta \psi_T^2 (\theta)}} .
    \label{eq:overlapIntegral}
\end{equation}
The modulus of the Poynting vector can be expanded as
\begin{equation}
    \delta |\boldsymbol{S}| = 2{\rm Re} \left[ \boldsymbol{S}^* \cdot \delta \boldsymbol{S} \right] ,
\end{equation}
where 
\begin{equation}
    \delta \boldsymbol{S} = \delta \boldsymbol{E} \times \boldsymbol{H}^* + \boldsymbol{E} \times \delta \boldsymbol{H}^* .
\end{equation}
Using these expressions to expand both the numerator and denominator in (\ref{eq:overlapIntegral}), we find an expression for how the overlap integral changes when the fields change by a small amount,
\begin{equation}
\begin{aligned}
    \delta I = \frac{1}{2 \sqrt{\int d\theta |\boldsymbol{S}(\theta)|^2 \int d\theta' \phi_{\rm T}^2 (\theta') }} \left[ \int \frac{d\theta}{|\boldsymbol{S} (\theta)|} {\rm Re} \left\{ \boldsymbol{S}^*(\theta) \cdot \left[ \delta \boldsymbol{E} \times \boldsymbol{H}^* + \boldsymbol{E} \times \delta \boldsymbol{H}^* \right] \right\} \phi_{\rm T} (\theta) \right. \\
    \left. - \frac{\int d\theta \phi_{\rm T} (\theta) |\boldsymbol{S} (\theta)|}{\int d\theta |\boldsymbol{S}(\theta)|^2} \int d\theta \ {\rm Re} \left\{ \boldsymbol{S}^*(\theta) \cdot \left[ \delta \boldsymbol{E} \times \boldsymbol{H}^* + \boldsymbol{E} \times \delta \boldsymbol{H}^* \right] \right\} \right] .
\end{aligned}
\end{equation}
Substituting into this the expressions of the variations of the fields (\ref{eq:deltaE}, \ref{eq:deltaH}), the gradient of the overlap integral can be written analytically. 

\subsection{Modulus of Poynting Vector in the Far--Field}

When designing the beaming device, shown in Figure \ref{fig:fig3}, the figure of merit is the modulus of the Poynting vector in the far--field when the structure is excited by a source of polarisation $\boldsymbol{p}_i$.
In this example, we have chosen the two circular polarisations.
For each polarisation, we write the figure of merit as is,
\begin{equation}
    F_i = |\boldsymbol{S}(\theta_i)|^2 .
\end{equation}
This means that for different source polarisations, power will be beamed into different directions.
The expansion of this follows the same procedure as the expansion of the modulus of the Poynting vector did in the overlap integral, giving the result,
\begin{align}
    \delta F_i &= 2{\rm Re} \left[ \boldsymbol{S}^*(\theta_i) \cdot \delta \boldsymbol{S}(\theta_i) \right] \\
    &= 2{\rm Re} \left[ \boldsymbol{S}^*(\theta_i) \cdot \left( \delta \boldsymbol{E}(\theta_i) \times \boldsymbol{H}^* (\theta_i) + \boldsymbol{E} (\theta_i) \times \delta \boldsymbol{H}^* (\theta_i) \right) \right].
\end{align}
Substituting into this the expressions for the field variations (\ref{eq:deltaE}, \ref{eq:deltaH}) gives analytic expressions for $\nabla_{\boldsymbol{r}_n} F_i$.

\subsection{Modulus of Electric Field}

For the `lensing' problem, demonstrated in Figure \ref{fig:fig4}, we seek to increase the modulus of the electric field at particular positions $\boldsymbol{r}_\star$,
\begin{equation}
    F = |\boldsymbol{E} (\boldsymbol{r}_\star)|^2 .
\end{equation}
The positions are different for each of the incidence directions.
This figure of merit can be easily expanded to find its gradient analytically,
\begin{align}
    \delta F &= 2 {\rm Re} \left[ \boldsymbol{E}^* (\boldsymbol{r}_\star) \cdot \delta \boldsymbol{E} (\boldsymbol{r}_\star) \right] , \\
    \nabla_{\boldsymbol{r}_n} F &= 2{\rm Re} \left[ \boldsymbol{E}^* (\boldsymbol{r}_\star) \cdot \left( \xi^2 \boldsymbol{G}(\boldsymbol{r}_\star, \boldsymbol{r}_n) \boldsymbol{\alpha}_E \nabla \boldsymbol{E} (\boldsymbol{r}_n) + i \xi \nabla \times \boldsymbol{G}(\boldsymbol{r}_\star, \boldsymbol{r}_n) \boldsymbol{\alpha}_H \nabla \boldsymbol{H} (\boldsymbol{r}_n) \right) \right] .
\end{align}

\section{Comparison with a Genetic Algorithm}

\begin{figure}
    \centering
    \includegraphics[width=0.8\linewidth]{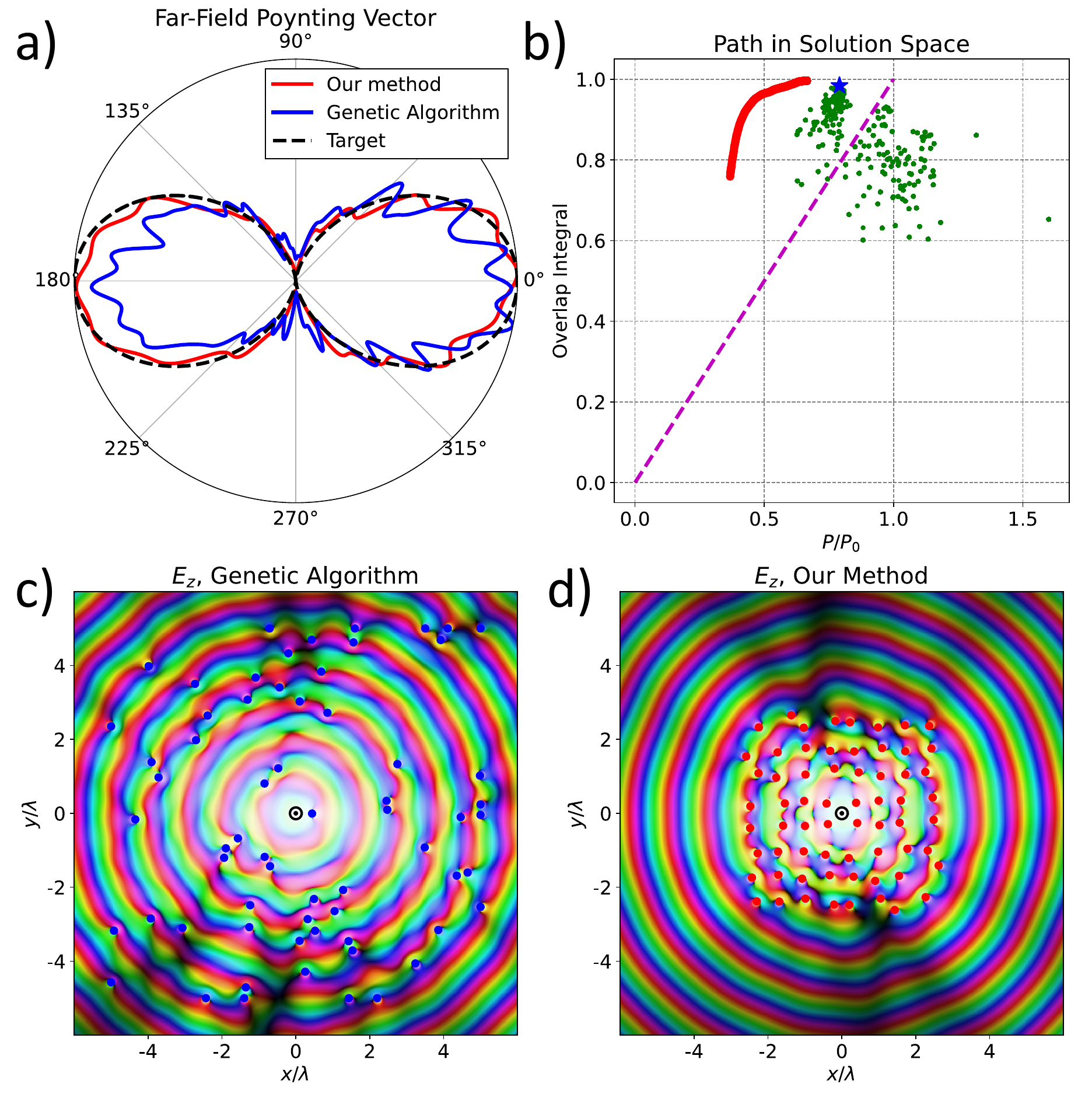}
    \caption{A comparison of the results of our optimisation and a genetic algorithm seeking to shape a far--field radiation pattern while also improving efficiency.
    The far--field radiation patterns are compared in a), and the solution space paths are shown in b).  The progress of our method is shown in red, and the progress of the genetic algorithm as green dots.  Each dot represents a single population member.  The final result of the genetic algorithm is shown as a blue star.
    The resulting structures are shown in c) and d).}
    \label{fig:go_comp}
\end{figure}

We compare the results of our optimisation for both power emission and directivity with the results of a genetic algorithm solving the same problem.
Using the differential evolution algorithm \cite{Storn1997}, with a population size of 20, and a maximum allowed iterations of 5000. 
The differential weight parameter is $F = 0.5$ and the crossover probability is CR = 0.7.
This genetic algorithm was run several times and the best solution selected.
The comparison between this result and the result of our local optimisation is shown in Figure \ref{fig:go_comp}.
The genetic algorithm produces a slightly higher power emission but a slightly lower value of overlap integral. 
From the scatter of the solutions generated by the genetic algorithm in solution space, shown in Figure \ref{fig:go_comp}, it is evident that the genetic algorithm explores more of the search space than our local optimisation.
However, due to the size of the search space for multi--functional problems, this does not provide much advantage.

\section{Validity of the Discrete Dipole Approximation}

To verify the validity of the discrete dipole approximation, we compare our results with full--wave solutions using a finite element method numerical solver (COMSOL Multiphysics) \cite{comsol}.

\begin{figure}
    \centering
    \includegraphics[width=\linewidth]{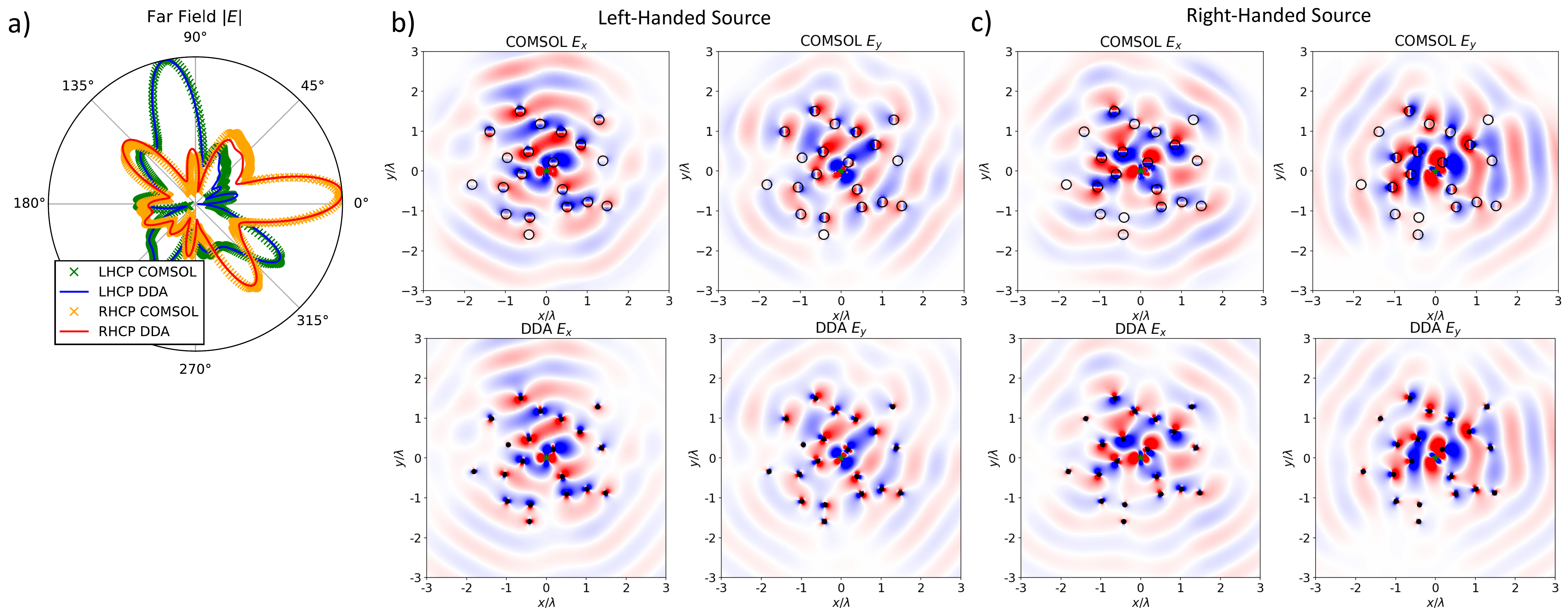}
    \caption{Comparison between our analytic results, based on the discrete dipole approximation, and full--wave simulations in COMSOL.  
    Considering only the 20 (of 64) scatterers nearest to the source of the device shown in Figure \ref{fig:fig3} of the main paper, we compare a) the far--fields and b),c) the near--fields.  
    For these scatterers, 65 nm radius silicon spheres, good agreement with the discrete dipole approximation is shown.}
    \label{fig:comsol_beaming}
\end{figure}

The beaming device shown in Figure \ref{fig:fig3} has been validated by considering the nearest 20 scatterers to the source, due to memory considerations.
For this reduced system, the comparison between the discrete dipole approximation and COMSOL is shown in Figure \ref{fig:comsol_beaming}.
The scatterers here are silicon spheres of radius 65 nm, for which the electric and magnetic polarisabilities can be found analytically.

\begin{figure}
    \centering
    \includegraphics[width=\linewidth]{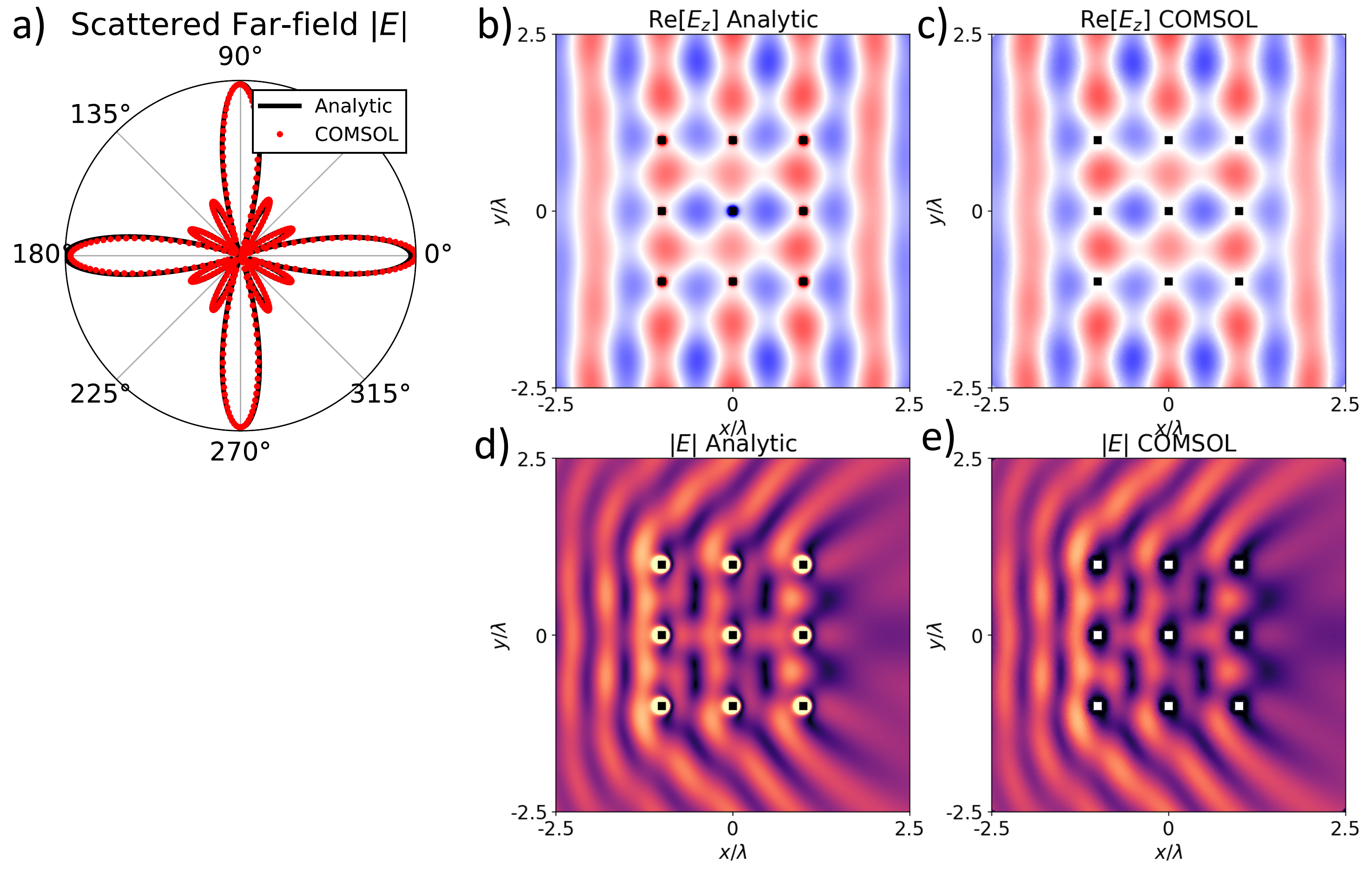}
    \caption{Comparison between the discrete dipole approximation and COMSOL for several metacubes under EM polarised plane wave incidence.
    Both a) the scattered field in the far--field and b)-e) the near--fields are shown.
    Excellent agreement between the analytics and full--wave simulations is found.}
    \label{fig:comsol_metacubes}
\end{figure}

Validation of the `lensing' device shown in Figure \ref{fig:fig4} of the main paper is shown in Figure \ref{fig:comsol_metacubes}.
We consider a TE plane wave incident upon a small number of metacubes.
Comparing the near and far fields in Figure \ref{fig:comsol_metacubes} the main difference is in the field at the location of the scatterers, where the discrete dipole approximation is not valid anyway.
The PEC boundary condition on the metal cubes in COMSOL ensures that the field inside the metacubes is zero.
However, in the analytics, the field on one of the scatterers is proportional to $\boldsymbol{G} (\boldsymbol{r}_n, \boldsymbol{r}_n)$.
The real part of this expression diverges, while the imaginary part remains finite.
The divergence of the real part is what cases the difference in the fields at the scatterer locations.

\bibliographystyle{naturemag}
\bibliography{refs}

\begin{thebibliography}{10}
\expandafter\ifx\csname url\endcsname\relax
  \def\url#1{\texttt{#1}}\fi
\expandafter\ifx\csname urlprefix\endcsname\relax\def\urlprefix{URL }\fi
\providecommand{\bibinfo}[2]{#2}
\providecommand{\eprint}[2][]{\url{#2}}

\bibitem{Munk2000}
\bibinfo{author}{Munk, B.~A.}
\newblock \emph{\bibinfo{title}{Frequency Selective Surfaces: Theory and
  Design}} (\bibinfo{publisher}{John Wiley and Sons}, \bibinfo{year}{2000}).

\bibitem{Leonhardt2006}
\bibinfo{author}{Leonhardt, U.} \& \bibinfo{author}{Philbin, T.~G.}
\newblock \bibinfo{title}{General relativity in electrical engineering}.
\newblock \emph{\bibinfo{journal}{New J. Phys.}} \textbf{\bibinfo{volume}{8}},
  \bibinfo{pages}{247} (\bibinfo{year}{2006}).
\newblock \urlprefix\url{https://doi.org/10.1088/1367-2630/8/10/247}.

\bibitem{Pendry2006}
\bibinfo{author}{Pendry, J.~B.}, \bibinfo{author}{Schurig, D.} \&
  \bibinfo{author}{Smith, D.~R.}
\newblock \bibinfo{title}{Controlling electromagnetic fields}.
\newblock \emph{\bibinfo{journal}{Science}} \textbf{\bibinfo{volume}{312}},
  \bibinfo{pages}{5781} (\bibinfo{year}{2006}).
\newblock \urlprefix\url{https://doi.org/10.1126/science.1125907}.

\bibitem{Galiffi2022}
\bibinfo{author}{Galiffi, E.} \emph{et~al.}
\newblock \bibinfo{title}{Photonics of time-varying media}.
\newblock \emph{\bibinfo{journal}{Adv. Photonics}}
  \textbf{\bibinfo{volume}{4}}, \bibinfo{pages}{014002} (\bibinfo{year}{2022}).
\newblock \urlprefix\url{https://doi.org/10.1117/1.AP.4.1.014002}.

\bibitem{Ptitcyn2022}
\bibinfo{author}{Ptitcyn, G.} \emph{et~al.}
\newblock \bibinfo{title}{Scattering from spheres made of time-varying and
  dispersive materials}  (\bibinfo{year}{2021}).
\newblock \urlprefix\url{https://arxiv.org/abs/2110.07195}.

\bibitem{Kadic2019}
\bibinfo{author}{Kadic, M.}, \bibinfo{author}{Milton, G.~W.},
  \bibinfo{author}{van Hecke, M.} \& \bibinfo{author}{Wegener, M.}
\newblock \bibinfo{title}{3d metamaterials}.
\newblock \emph{\bibinfo{journal}{Nat. Rev. Phys.}}
  \textbf{\bibinfo{volume}{1}}, \bibinfo{pages}{198--210}
  (\bibinfo{year}{2019}).
\newblock \urlprefix\url{https://doi.org/10.1038/s42254-018-0018-y}.

\bibitem{Pande2020}
\bibinfo{author}{Pande, D.}, \bibinfo{author}{Gollub, J.},
  \bibinfo{author}{Zecca, R.}, \bibinfo{author}{Marks, D.~L.} \&
  \bibinfo{author}{Smith, D.~R.}
\newblock \bibinfo{title}{Symphotic multiplexing medium at microwave
  frequencies}.
\newblock \emph{\bibinfo{journal}{Phys. Rev. Applied}}
  \textbf{\bibinfo{volume}{13}}, \bibinfo{pages}{024033}
  (\bibinfo{year}{2020}).
\newblock \urlprefix\url{https://doi.org/10.1103/PhysRevApplied.13.024033}.

\bibitem{Frellsen2016}
\bibinfo{author}{Frellsen, L.~F.}, \bibinfo{author}{Ding, Y.},
  \bibinfo{author}{Sigmund, O.} \& \bibinfo{author}{Frandsen, L.~H.}
\newblock \bibinfo{title}{Topology optimized mode multiplexing in
  silicon-on-insulator photonic wire waveguides}.
\newblock \emph{\bibinfo{journal}{Opt. Express}} \textbf{\bibinfo{volume}{24}},
  \bibinfo{pages}{16866--16873} (\bibinfo{year}{2016}).
\newblock \urlprefix\url{https://doi.org/10.1364/OE.24.016866}.

\bibitem{Piggott2015}
\bibinfo{author}{Piggott, A.~Y.} \emph{et~al.}
\newblock \bibinfo{title}{Inverse design and demonstration of a compact and
  broadband on-chip wavelength demultiplexer}.
\newblock \emph{\bibinfo{journal}{Nature Photon.}}
  \textbf{\bibinfo{volume}{9}}, \bibinfo{pages}{374--377}
  (\bibinfo{year}{2015}).
\newblock \urlprefix\url{https://doi.org/10.1038/nphoton.2015.69}.

\bibitem{Li2019}
\bibinfo{author}{Li, S.-Q.} \emph{et~al.}
\newblock \bibinfo{title}{Phase-only transmissive spatial light modulator based
  on tunable dielectric metasurface}.
\newblock \emph{\bibinfo{journal}{Science}} \textbf{\bibinfo{volume}{364}},
  \bibinfo{pages}{1087--1090} (\bibinfo{year}{2019}).
\newblock \urlprefix\url{https://doi.org/10.1126/science.aaw6747}.

\bibitem{Wiecha2017}
\bibinfo{author}{Wiecha, P.~R.} \emph{et~al.}
\newblock \bibinfo{title}{Evolutionary multi-objective optimization of colour
  pixels based on dielectric nanoantennas}.
\newblock \emph{\bibinfo{journal}{Nature Nanotech.}}
  \textbf{\bibinfo{volume}{12}}, \bibinfo{pages}{163--169}
  (\bibinfo{year}{2017}).
\newblock \urlprefix\url{https://doi.org/10.1038/nnano.2016.224}.

\bibitem{Yeung2020}
\bibinfo{author}{Yeung, C.} \emph{et~al.}
\newblock \bibinfo{title}{Elucidating the behavior of nanophotonic structures
  through explainable machine learning algorithms}.
\newblock \emph{\bibinfo{journal}{ACS Photonics}} \textbf{\bibinfo{volume}{7}},
  \bibinfo{pages}{2309--2318} (\bibinfo{year}{2020}).
\newblock \urlprefix\url{https://doi.org/10.1021/acsphotonics.0c01067}.

\bibitem{Stratton2007}
\bibinfo{author}{Stratton, J.~A.}
\newblock \emph{\bibinfo{title}{Electromagnetic Theory}}
  (\bibinfo{publisher}{John Wiley and Sons}, \bibinfo{year}{2007}).

\bibitem{Miller2021}
\bibinfo{author}{Shim, H.}, \bibinfo{author}{Kuang, Z.}, \bibinfo{author}{Lin,
  Z.} \& \bibinfo{author}{Miller, O.~D.}
\newblock \bibinfo{title}{Fundamental limits to multi-functional and tunable
  nanophotonic response}.
\newblock \emph{\bibinfo{journal}{arXiv:2112.10816v1}}  (\bibinfo{year}{2021}).
\newblock \urlprefix\url{https://arxiv.org/pdf/2112.10816}.

\bibitem{Molesky2018}
\bibinfo{author}{Molesky, S.} \emph{et~al.}
\newblock \bibinfo{title}{Inverse design in nanophotonics}.
\newblock \emph{\bibinfo{journal}{Nature Photon.}}
  \textbf{\bibinfo{volume}{12}}, \bibinfo{pages}{659--670}
  (\bibinfo{year}{2018}).
\newblock \urlprefix\url{https://doi.org/10.1038/s41566-018-0246-9}.

\bibitem{Gerchberg1972}
\bibinfo{author}{Gerchberg, R.~W.} \& \bibinfo{author}{Saxton, W.~O.}
\newblock \bibinfo{title}{A practical algorithm for the determination of the
  phase from image and diffraction plane pictures}.
\newblock \emph{\bibinfo{journal}{Optik}} \textbf{\bibinfo{volume}{35}},
  \bibinfo{pages}{237--246} (\bibinfo{year}{1972}).
\newblock
  \urlprefix\url{http://www.u.arizona.edu/~ppoon/GerchbergandSaxton1972.pdf}.

\bibitem{Liu2021}
\bibinfo{author}{Liu, M.} \emph{et~al.}
\newblock \bibinfo{title}{Multifunctional metasurfaces enabled by simultaneous
  and independent control of phase and amplitude for orthogonal polarization
  states}.
\newblock \emph{\bibinfo{journal}{Light Sci. Appl.}}
  \textbf{\bibinfo{volume}{10}}, \bibinfo{pages}{107} (\bibinfo{year}{2021}).
\newblock \urlprefix\url{https://doi.org/10.1038/s41377-021-00552-3}.

\bibitem{Bendsoe2003}
\bibinfo{author}{Bendsøe, M.~P.} \& \bibinfo{author}{Sigmund, O.}
\newblock \emph{\bibinfo{title}{Topology optimization: Theory, methods and
  applications}} (\bibinfo{publisher}{Springer}, \bibinfo{year}{2003}).

\bibitem{Lalau-Keraly2013}
\bibinfo{author}{Lalau-Keraly, C.~M.}, \bibinfo{author}{Bhargava, S.},
  \bibinfo{author}{Miller, O.~D.} \& \bibinfo{author}{Yablonovitch, E.}
\newblock \bibinfo{title}{Adjoint shape optimization applied to electromagnetic
  design}.
\newblock \emph{\bibinfo{journal}{Opt. Express}} \textbf{\bibinfo{volume}{21}},
  \bibinfo{pages}{21693--21701} (\bibinfo{year}{2013}).
\newblock \urlprefix\url{https://doi.org/10.1364/OE.21.021693}.

\bibitem{Purcell1973}
\bibinfo{author}{Purcell, E.~M.} \& \bibinfo{author}{Pennypacker, C.~R.}
\newblock \bibinfo{title}{Scattering and absorption of light by nonspherical
  dielectric grains}.
\newblock \emph{\bibinfo{journal}{Astrophys. J.}}
  \textbf{\bibinfo{volume}{186}}, \bibinfo{pages}{705} (\bibinfo{year}{1973}).
\newblock \urlprefix\url{https://doi.org/10.1086/152538}.

\bibitem{Powell2020}
\bibinfo{author}{Powell, A.~W.} \emph{et~al.}
\newblock \bibinfo{title}{Strong, omnidirectional radar backscatter from
  subwavelength, 3d printed metacubes}.
\newblock \emph{\bibinfo{journal}{IET Microw. Antennas Propag.}}
  \textbf{\bibinfo{volume}{14}}, \bibinfo{pages}{1862--1868}
  (\bibinfo{year}{2020}).

\bibitem{Schwinger1950}
\bibinfo{author}{Levine, H.} \& \bibinfo{author}{Schwinger, J.}
\newblock \bibinfo{title}{On the theory of electromagnetic wave diffraction by
  an aperture in an infinite plane conducting screen}.
\newblock \emph{\bibinfo{journal}{Commun. Pure Appl. Math}}
  \textbf{\bibinfo{volume}{3}}, \bibinfo{pages}{355--391}
  (\bibinfo{year}{1950}).

\bibitem{Tai1993}
\bibinfo{author}{Tai, C.-T.}
\newblock \emph{\bibinfo{title}{Dyadic Greens Functions in Electromagnetic
  Theory}} (\bibinfo{publisher}{IEEE Press}, \bibinfo{address}{New York},
  \bibinfo{year}{1993 (second edition)}).

\bibitem{Foldy1945}
\bibinfo{author}{Foldy, L.~L.}
\newblock \bibinfo{title}{{The Multiple Scattering of Waves. I. General Theory
  of Isotropic Scattering by Randomly Distributed Scatterers}}.
\newblock \emph{\bibinfo{journal}{Physical Review}}
  \textbf{\bibinfo{volume}{67}}, \bibinfo{pages}{107} (\bibinfo{year}{1945}).
\newblock \urlprefix\url{https://doi.org/10.1103/PhysRev.67.107}.

\bibitem{Capers2021}
\bibinfo{author}{J.~R.~Capers, A. P.~H., S. J.~Boyes} \&
  \bibinfo{author}{Horsley, S. A.~R.}
\newblock \bibinfo{title}{Designing the collective non-local responses of
  metasurfaces}.
\newblock \emph{\bibinfo{journal}{Communications Physics}}
  \textbf{\bibinfo{volume}{4}}, \bibinfo{pages}{209} (\bibinfo{year}{2021}).
\newblock \urlprefix\url{https://doi.org/10.1038/s42005-021-00713-1}.

\bibitem{Shwartz2014}
\bibinfo{author}{Shalev-Shwartz, S.} \& \bibinfo{author}{Ben-David, S.}
\newblock \emph{\bibinfo{title}{Understanding Machine Learning: From Theory to
  Algorithms}} (\bibinfo{publisher}{Cambridge University Press},
  \bibinfo{year}{2014}).

\bibitem{Hwang1979}
\bibinfo{author}{Hwang, C.-L.} \& \bibinfo{author}{Masud, A.~S.}
\newblock \emph{\bibinfo{title}{Multiple Objective Decision Making: Methods and
  Applications}} (\bibinfo{publisher}{Springer--Verlag}, \bibinfo{year}{1979}).

\bibitem{comsol}
\bibinfo{title}{Comsol multiphysics v. 6.0, \url{http://www.comsol.com}}.

\bibitem{Arango2013}
\bibinfo{author}{Arango, F.~B.} \& \bibinfo{author}{Koenderink, A.~F.}
\newblock \bibinfo{title}{Polarizability tensor retrieval for magnetic and
  plasmonic antenna design}.
\newblock \emph{\bibinfo{journal}{New J. Phys.}} \textbf{\bibinfo{volume}{15}},
  \bibinfo{pages}{073023} (\bibinfo{year}{2013}).

\bibitem{Liu2016}
\bibinfo{author}{Liu, X.-X.}, \bibinfo{author}{Zhao, Y.} \&
  \bibinfo{author}{Alù, A.}
\newblock \bibinfo{title}{Polarizability tensor retrieval for subwavelength
  particles of arbitrary shape}.
\newblock \emph{\bibinfo{journal}{IEEE Trans. Antennas Propag.}}
  \textbf{\bibinfo{volume}{64}}, \bibinfo{pages}{2301--2310}
  (\bibinfo{year}{2016}).

\bibitem{Raab2005}
\bibinfo{author}{Raab, R.~E.} \& \bibinfo{author}{de~Lange, O.~L.}
\newblock \emph{\bibinfo{title}{Multipole Theory in Electromagnetism}}
  (\bibinfo{publisher}{Oxford University Press}, \bibinfo{address}{Oxford},
  \bibinfo{year}{2005}).

\bibitem{Evlyukhin2011}
\bibinfo{author}{Evlyukhin, A.~B.}, \bibinfo{author}{Reinhardt, C.} \&
  \bibinfo{author}{Chichkov, B.~N.}
\newblock \bibinfo{title}{Multipole light scattering by nonspherical
  nanoparticles in the discrete dipole approximation}.
\newblock \emph{\bibinfo{journal}{Phys. Rev. B}} \textbf{\bibinfo{volume}{84}},
  \bibinfo{pages}{235429} (\bibinfo{year}{2011}).

\bibitem{Evlyukhin2013}
\bibinfo{author}{Evlyukhin, A.~B.}, \bibinfo{author}{Reinhard, C.},
  \bibinfo{author}{Evlyukhin, E.} \& \bibinfo{author}{Chichkov, B.~N.}
\newblock \bibinfo{title}{Multipole analysis of light scattering by
  arbitrary-shaped nanoparticles on a plane surface}.
\newblock \emph{\bibinfo{journal}{J. Opt. Soc. Am. B}}
  \textbf{\bibinfo{volume}{30}}, \bibinfo{pages}{2589--2598}
  (\bibinfo{year}{2013}).

\bibitem{Storn1997}
\bibinfo{author}{Storn, R.} \& \bibinfo{author}{Price, K.}
\newblock \bibinfo{title}{Differential evolution – a simple and efficient
  heuristic for global optimization over continuous spaces}.
\newblock \emph{\bibinfo{journal}{Journal of Global Optimization}}
  \textbf{\bibinfo{volume}{11}}, \bibinfo{pages}{314--359}
  (\bibinfo{year}{1997}).
\newblock \urlprefix\url{https://doi.org/10.1023/A:1008202821328}.

\end{thebibliography}

\end{document}